\documentstyle[12pt]{article}
\begin{document}
\begin{center}
\section*{\bf Gamma Ray Burst as Vacuum Discharge of Super-Schwinger Electric
Fields}

\vspace{5mm}
R. Lieu$^1$, Y. Takahashi$^1$ and T.W.B. Kibble$^2$

\vspace{1.5mm}

$^1$ Department of Physics, University of Alabama, Huntsville, AL 35899. \\
({\tt lieur@cspar.uah.edu, yoshi@cosmic.uah.edu}) \\
$^2$ Blackett Laboratory, Imperial College, London SW 7 2BZ. \\
({\tt t.kibble@ic.ac.uk}) \\

\end{center}

\vspace{2mm}

{\it A theory is proposed to explain with
simplicity the basic observed properties of a Gamma Ray Burst (GRB).
It employs a well-known result of Schwinger, that static electric fields
in excess of a critical value are unstable to pair creation, and
catastrophically produces a thermal plasma at temperatures $\leq$ 0.5 MeV.
By using observational values for the energy
and volume of the source, it is shown that
the radiation pressure of an expanding GRB `fireball'
leads to the formation of a Schwinger critical field at the ambient
medium immediately outside the `fireball'.  This
naturally provides a runaway solution which is inevitable, and which must
involve a burst of gamma radiation in the core of the observed
energy range and in an
optically thin environment.  The observed burst duration of
1 -- 10 seconds is also a straightforward consequence of the theory.}

\vspace{2.5mm}

It has long been recognized (Schwinger 1951) that when a static
electric field exceeds a critical value 
$E = m_e^2 c^3/e \hbar = E_c$, corresponding to an 
electron acceleration of
a$_c \sim$ 2.4 $\times$ 10$^{31}$ cm s$^{-2}$, it is unstable with
respect to pair production.  Under such a circumstance, a virtual
pair of electron and positron is accelerated
in opposite directions
by the electric field to the speed of light within a Compton
wavelength.  This greatly reduces the probability of annihilation:
real pair production may then take place at the energy expense of the
field.  The process is analogous to a lightning discharge, except that
it happens in vacuum.
The rate of pair production per unit volume is given by the formula
Schwinger derived:
\begin{eqnarray}
\frac{d^2N}{dVdt} = \frac{\alpha^2 E^2}{\pi^2 \hbar}
\sum_{n=1}^{\infty} n^{-2} exp 
\left(\frac{-n \pi E_c}{E} \right) \nonumber
\end{eqnarray}
where $\alpha$ is the fine
structure constant and $N$ is the number of pairs created.  Using
this formula, we find that
for $E \sim E_c$ the `vacuum breakdown' causes the field to dissipate
its energy in a timescale of $\sim$ 10$^{-16}$ seconds, resulting
in a mixture of gamma rays and pair plasma at temperature $\leq$ 0.5 MeV.
Owing to the strength of $E_c$, its instability has not been realized
in the laboratory.
This paper outlines a theory of GRB based on
the Schwinger mechanism (SM).  

The first detection of delayed
GRB afterglows at optical wavelength (van
Paradijs et al 1997) quickly led to a settlement of the question concerning the
distance scale: GRB must be of cosmological origin.  Furthermore,
the high redshift association
($0.8 < z < 2.0$, see
Bond 1997, Frail and Kulkarni 1997) and the presence of deep and
rapid temporal modulations in the GRB flux (Bhat et al 
1992) render it by now
a generally accepted fact that the intrinsic energy of a GRB is
$\geq$ 10$^{52}$ ergs and the size of the burst region (the `fireball') is
$\leq$ 3 $\times$ 10$^{7}$ cm.  

The data immediately pose extreme
difficulties in the search for a viable GRB emission mechanism:
even at gamma ray energies the Thomson
optical depth at such high densities is $\gg$ 1, implying that 
radiation cannot leave the source unless the
latter has expanded sufficiently to
become optically thin.  By then, however, the source size will be
incompatible
with that imposed by the observed time variability.  Furthermore,
it is unclear how a relativistic expansion model leads to the
considerable observable
consistencies in the basic properties of GRBs, especially given the wide
range of Lorentz factor dependence of the various parameters (from $\gamma$
to $\gamma^6$, see Meszaros and Rees 1993).  Thus, e.g., why should the
emission of a high $\gamma$ system be invariably peaked at around the pair
production energy ?  We emphasize that a relativistic bulk outflow is
{\it not} an inevitable fate of the `fireball'.  This is only so if the
system has a very small baryonic fraction (Meszaros and Rees 1993).
If on the other hand the `fireball' contains 1 M$_\odot$ worth of baryonic
matter (as is reasonably the case, since the `host' of the
`fireball' is likely to be $\geq$ 1 M$_\odot$ massive, see below)
the gravitational potential would be sufficient to bind a system of
protons, electrons, positrons and radiation
in thermal equilibrium
at kT $\leq$ 10 MeV within the `fireball' radius.  Although the radiation
pressure is super-Eddington the system cannot overcome gravitation, the
result is oscillations of
the `envelope' with the dynamic timescale of $\sim$ 10$^{-3}$ s, which may
be responsible for the observed variability.

We propose here an alternative model of GRB which can explain its basic
properties without invoking extreme assumptions.  We are not primarily
concerned with the origin of the energy in the `fireball', except to say
that we presume a very rapid energy release over timescales $\sim$ 
10$^{-3}$ s,
such as by neutron star merging or stellar collapse
(Paczynski 1993), all
of which involves $\geq$ 1 M$_\odot$ object(s) at the center of the
system.
Since a quantum must diffuse very slowly out of the `fireball', the
correct way of estimating the energy density is to divide the
total energy of the GRB by the source volume.  Using the numbers
quoted above, one obtains a value of 
$U_{GRB} \sim$ 10$^{29}$ ergs cm$^{-3}$, which corresponds to a black
body temperature of kT $\sim$ 5 MeV.  Since the dense environment of the
`fireball' necessitates very frequent collisions, baryons and radiation
cannot avoid the condition of thermal equilibrium.  This implies that
electrons and protons also have kT $\sim$ 5 MeV, and the system is
gravitationally bound as argued before.

To appreciate the significance of SM in the understanding of GRB, one
must  first realize that the radiation pressure exerted by an
expanding `fireball'
of energy density $U_{GRB}$ on the
ambient matter immediately outside it is enormous, {\it sufficient
to cause a vacuum breakdown}.  This is demonstrated
by calculating the electron acceleration due to radiation:
a $= \sigma_{Th} U_{GRB}/m_e \sim$ 7 $\times$ 10$^{31}$ cm s$^{-2} >$ a$_c$.
Since proton acceleration is reduced by a factor of $(m_e/m_p)^3$,
momentarily the electrons advance radially outwards while the protons
remain effectively stationary.
This means a static electric  field
$E > E_c$ must exist
between the two types of particles, because the electrons move
with a $>$ a$_c$ with respect to the protons.  Such a field will
discharge by the SM, and will convert the surface flow energy of
the `fireball' into thermal energy at kT $\leq$ 0.5 MeV
{\it in an optically thin region} just outside it.  The
discharge is self-sustaining in that it will propagate into the `fireball'
and `extract' the bulk of its energy into an avalanche of electron-
positron pairs.  The ensuing annihilation produces gamma 
rays which {\it can}
escape.
The result is the
GRB that we observe.

Another basic property of GRB which should be explained is the
observed burst duration, typically 1 -- 10 seconds.
{\it This is a natural consequence of the theory}.
The reasoning is as follows.  A static electric field of strength
$E \geq E_c$ must have length $\geq$ an electron Compton wavelength
$\sim 2.5 \times 10^{-10}$ cm
before it can discharge.  At shorter lengths there will not be
sufficient distance for an electron to separate from a positron - a
necessary condition for the formation of real pairs.  Thus the
`fireball' radiation pressure must push out the ambient plasma
selectively to produce a discharge layer of thickness
$\sim 2.5 \times 10^{-10}$ cm, the SM will then commence, with a
discharge time of 10$^{-16}$ s, as given above.
The ratio of these two numbers implies that
the discharge propagates into the `fireball' at a speed
of 2.5 $\times$ 10$^6$ cm per sec.  Since the `fireball' has size
$\sim$ 3 $\times$ 10$^7$ cm, it will be consumed in a timescale
of $\sim$ 1 -- 10 seconds, in excellent agreement with the
observed duration of a GRB.  This slow, inward moving
discharge layer provides ample time for the gamma rays 
and pairs already created
outside it by the 
SM to escape from the optically thick `fireball' (none of
the created quanta is gravitationally bound).  Thus an
observable GRB event
{\it can} emerge from a compact region.

In conclusion, we propose a theory which explains with
simplicity the basic observed properties of a GRB.  Firstly,
the significance of an intrinsic energy $\sim$ 10$^{52}$ ergs
and a size $\sim$ 3 $\times$ 10$^7$ cm may now be understood:
if a major perturbation occurs to these two numbers the radiation
pressure of the `fireball' on the ambient medium will either generate
sub-Schwinger electric fields, resulting in no discharge, or
super-Schwinger fields with discharge having happened in advance
to reduce the energy density.  Secondly, the theory naturally predicts
a catastrophic runaway in the form of a burst of gamma rays: the duration
of a GRB is 1 -- 10 seconds due to the speed of the propagation of the
discharge, and the radiation is gamma rays due to the post-discharge
temperature.  The usual problem of source optical thickness is not
a difficulty here.

The theory may be relevant to supernova explosions also, since
the core electromagnetic energy density is $\sim$ that of a GRB.  If
correct, these dramatic astronomical events will become live witnesses
of Schwinger's great prediction 
- a prediction which has hitherto not
been verified in terrestrial environments.

We thank Martin Rees, Jan van Paradijs, Walter Lewin and Gordon
Emslie for helpful discussions.

\vspace{2mm}

\noindent
{\bf References}

\noindent
Bhat, P.N. et al, 1992, Nature, 359, 217. \\
\noindent
Bond, H.E. 1997, IAU Circ 6654. \\
\noindent
Frail, D.A. \& Kulkarni, S.R. 1997, IAU Circ 6662. \\
\noindent
Meszaros, P. \& Rees, M.J., 1993, ApJ 405, 278. \\
\noindent
Paczynski, B. 1993, in `Compton Gamma Ray Observatory', ed.
Friedlander et al (New York: AIP), 981. \\
\noindent
Schwinger, J. 1951, Phys. Rev. 82, 664. \\
\noindent
van Paradijs, J. et al 1997, Nature, 386, 686. \\

\end{document}